\documentclass{aa}  
\usepackage{graphicx}
\usepackage{float}
\usepackage{natbib}
\usepackage{epstopdf}
\usepackage{lscape}
\usepackage{amsmath}
\usepackage{url}
\usepackage{soul}
\usepackage{ulem}
\usepackage{cancel}
\setstcolor{red}
\usepackage[utf8]{inputenc}
\usepackage[colorlinks=true,urlcolor=blue,linkcolor=blue,anchorcolor=blue,citecolor=blue]{hyperref}
\usepackage{appendix}
\usepackage{txfonts}
\begin{document}

   \title{Radial velocity variability fractions of different types of hot subdwarf stars}

   \author{Ruijie He\inst{1,2}, Xiangcun Meng\inst{1,3,4}, Zhenxin Lei\inst{5,6}, Huahui Yan\inst{7}, Shunyi Lan\inst{1,2}}

   \authorrunning{Ruijie He et al.}
   
   \institute{Yunnan Observatories, Chinese Academy of Sciences, Kunming 650011, PR China\\e-mail:heruijie@ynao.ac.cn, xiangcunmeng@ynao.ac.cn
  \and University of the Chinese Academy of Sciences, Beijing 100049, PR China
  \and Key Laboratory for the Structure and Evolution of Celestial Objects, Chinese Academy of Sciences, Kunming 650216, PR China
  \and International Centre of Supernovae, Yunnan Key Laboratory, Kunming 650216, PR China           
  \and Key Laboratory of Stars and Interstellar Medium, Xiangtan University, Xiangtan 411105, PR China
  \and Physics Department, Xiangtan University, Xiangtan 411105, PR China
  \and Shandong Provincial Key Laboratory of Optical Astronomy and Solar-Terrestrial Environment, School of Space Science and Physics, Shandong University at Weihai, Weihai 264209, PR China}
  \abstract
   {It is generally thought that hot subdwarfs are helium-core- or helium-shell-burning objects with extremely thin hydrogen envelopes and that binary interactions are always needed in their formation. Different types of hot subdwarfs may have different origins, which will cause them to present different radial velocity (RV) variability properties.}
   {We plan to study the RV-variability fractions of different types of hot subdwarfs, as well as their distributions in the $T_{\rm eff}$ -- $\log g$ diagram. This provides insights into the formation of hot subdwarfs.
   }
   {The cross-correlation function method was adopted to measure RV variations in 434 hot subdwarfs based on spectra obtained with the Large Sky Area Multi-Object Fiber Spectroscopic Telescope (LAMOST). Light curves from the Transiting Exoplanet Survey were also used to search for short-period binary hot subdwarfs.}
   {Only 6 $\pm$ 4\% of our single-lined He-rich hot subdwarfs that only show spectroscopic features of hot subdwarfs are found to be RV variable, which is lower than the fraction of single-lined He-poor sdB stars (31 $\pm$ 3\%). Single-lined sdB stars with effective temperatures ($T_{\rm eff}$) $\sim$ 25,000 -- 33,000 K show an RV-variability fraction of 34 $\pm$ 5\%, while lower RV-variability fractions are observed for single-lined sdB stars cooler than about 25,000 K (11 $\pm$ 4\%), single-lined sdB/OB stars with $T_{\rm eff}$ $\sim$ 33,000 -- 40,000 K and surface gravities $\sim$ 5.7 -- 6.0 (13 $\pm$ 3\%), as well as single-lined sdO/B stars with $T_{\rm eff}$ $\sim$ 45,000 -- 70,000 K (10 $\pm$ 7\%). Single-lined hot subdwarfs with $T_{\rm eff}$ $\sim$ 35,000 – 45,000 K located above the extreme horizontal branch (EHB) show a similar RV-variability fraction of 34 $\pm$ 9\% as single-lined sdB stars at about 25,000 – 33,000 K. The largest RV-variability fraction of 51 $\pm$ 8\% is found in single-lined hot subdwarfs below the canonical EHB. The detected RV-variability fraction of our composite hot subdwarfs with an infrared excess in their spectral energy distributions is 9 $\pm$ 3\%, which is lower than that fraction of single-lined hot subdwarfs. Since the average RV uncertainty we measured in the LAMOST spectra is about 7.0 km/s, the lower detected RV-variability fraction for composite hot subdwarfs is expected because the RV amplitudes associated with long-period systems are lower.} 
   {The results here are generally consistent with the canonical binary evolution channels for forming hot subdwarfs. Most single-lined He-rich hot subdwarfs may form through merger channels, while the stable Roche-lobe overflow channel could play an important role in the formation of composite hot subdwarfs. Single-lined hot subdwarfs with $T_{\rm eff}$ $\sim$ 35,000 – 45,000 K located above the EHB may have an evolutionary connection to the sdB stars at about 25,000 -- 33,000 K. The different detected RV-variability fractions for the different subclasses of single-lined hot subdwarfs indicate that their formation channels may differ.}

   \maketitle
   

\section{Introduction}

   Most hot subdwarf stars are thought to be in the helium-core (He-core) or He-shell-burning stage. They are located at the extreme blue end of the horizontal branch (HB) in the Hertzsprung-Russell (HR) diagram and are also known as extreme horizontal branch (EHB) stars. The masses of canonical hot subdwarfs are about 0.5 $M_\odot$ and their hydrogen envelopes are extremely thin (< 0.02 $M_\odot$) \citep{Heber.etal.1986}. Their effective temperatures ($T_{\rm eff}$) range between about 20,000 -- 80,000 K \citep[see][for a detailed review]{Heber.etal.2009,Heber.etal.2016}. Hot subdwarfs play very important roles in many fields in astrophysics. They may be the main origin of the ultraviolet upturn in elliptical galaxies \citep{Han.etal.2007}. Short-period binaries that are composed of a hot subdwarf and a massive white dwarf (WD) are good candidates for the progenitors of type Ia supernovae (SNe Ia) \citep{Maxted.etal.2000,Geier.etal.2007,Geier.etal.2013,Wang.etal.2010,Pelisoli.etal.2021}. Additionally, some short-period binary hot subdwarfs with WD companions are also important gravitational wave (GW) sources for future studies \citep{Kupfer.etal.2018,Lin.etal.2024}.
   
   About one-third of the hot subdwarfs are discovered in short-period, single-lined binaries \citep{Maxted.etal.2001,Napiwotzki.etal.2004,Copperwheat.etal.2011,Kawka.etal.2015,Geier.etal.2022}, which only present the spectroscopic features of hot subdwarfs. The orbital periods of these systems range from about one hour to several days, and their companions are mainly M-type main-sequence (dM), WD, or brown dwarf (BD) stars \citep[e.g.,][]{Kupfer.etal.2015,Schaffenroth.etal.2022}. Another distinct type of stars from single-lined objects are composite hot subdwarfs. Because these objects usually have F-, G-, or K-type companions, an infrared (IR) excess can be detected in them \citep{Heber.etal.2018,Solano.etal.2022}. To date, all of the composite hot subdwarf binaries with orbital solutions were found to be long-period systems with periods from about 400 to 1,500 days \citep{Vos.etal.2017,Vos.etal.2018,Vos.etal.2019}.

   In order to form hot subdwarfs, most of the hydrogen envelopes of their progenitors need to be lost before the He-core-burning phase. This mass loss is very difficult to achieve in the context of single-star evolution, and binary evolution is therefore usually proposed \citep{Heber.etal.2009,Heber.etal.2016}. The detailed binary population synthesis (BPS) calculations from \citet{Han.etal.2002,Han.etal.2003} indicated that stable Roche-lobe overflow (RLOF), common-envelope (CE) ejection, and double He-WD mergers are the main formation channels for hot subdwarfs. Long-period hot subdwarf binaries result from the stable RLOF channel \citep{Han.etal.2002,Han.etal.2003,Vos.etal.2020,Chen.etal.2013,Gotberg.etal.2018}. Short-period binary hot subdwarfs can only form through the CE ejection channel \citep{Han.etal.2002,Han.etal.2003,Xiong.etal.2017,Ge.etal.2024,Ge.etal.2022}. The merger of double He-WDs in a short-period binary only produces single hot subdwarfs \citep{Han.etal.2002,Han.etal.2003,Webbink.etal.1984,Zhang.etal.2012,Hall.etal.2016,Schwab.etal.2018}, and the merger products are typically considered to be He-rich hot subdwarfs \citep{Zhang.etal.2012,Schwab.etal.2018}. Additionally, some other models were proposed. When the merger occurs in a binary containing a red giant star (RGB) and a dM or BD companion, or even a high-mass planet during the CE phase, a rapidly rotating HB star may be produced \citep{Kramer.etal.2020,Soker.etal.1998,Politano.etal.2008}. The centrifugal force for rapid rotation may enhance the mass loss from the HB star, and an sdB star may form. \citet{Justham.etal.2011} suggested that the merger of a hybrid CO-He WD and a He-WD can form He-sdO stars. Moreover, He-WDs that merge with their dM companions may initially form intermediate He-rich (iHe-rich) sdOB stars. Subsequently, the element diffusion processes may change them into He-poor sdOB stars \citep{Zhang.etal.2017}. Some single iHe-rich sdB stars might be surviving companions of SNe Ia from the WD+MS channel \citep[e.g.,][]{Meng.etal.2019,Meng.etal.2021}.
   
   Studies of radial velocity (RV) variable stars revealed that a substantial number of hot subdwarfs exist in short-period binaries. Furthermore, the RV-variability fractions differ among different types of hot subdwarfs, indicating that they may have different origins \citep[e.g.,][]{Maxted.etal.2001,Napiwotzki.etal.2004,Copperwheat.etal.2011,Kawka.etal.2015,Geier.etal.2022}. \citet{Maxted.etal.2001} observed 36 EHB stars over a time span of 11 nights to determine their RV variations. Among the five single-lined post-EHB stars and five composite EHB stars, only one single-lined post-EHB star was detected with significant RV variations. In contrast, a high RV-variability fraction of 68\% was observed in their 26 single-lined EHB stars. Subsequently, by using the randomly observed spectra from the ESO Supernova Ia Progenitor Survey \citep[SPY,][]{Napiwotzki.etal.2001}, \citet{Napiwotzki.etal.2004} found that the RV-variability fractions for 46 single-lined sdB and 23 single-lined He-sdO stars were 39\% and 4\%, respectively. The long-term monitoring survey conducted by \citet{Morales.etal.2003} and \citet{Copperwheat.etal.2011} revealed that 64 out of 125 (51\%) single-lined sdB stars and 5 out of 32 (16\%) composite sdB stars exhibit RV variability. In summary, most He-rich hot subdwarfs usually do not show significant RV variations. Most composite hot subdwarfs are also not detected with significant RV variations based on the limited RV accuracy achieved by low- or medium-resolution spectra. These two types of stars are quite different from the single-lined He-poor sdB stars. 

   Recently, \citet{Geier.etal.2022} have measured the RV variations for 646 single-lined hot subdwarfs by using the multi-epoch spectra from the Sloan Digital Sky Survey Data Release 12 \citep[SDSS DR12,][]{Alam.etal.2015} and the Large Sky Area Multi-Object Fiber Spectroscopic Telescope Data Release 5 \citep[LAMOST DR5,][]{Cui.etal.2012}. Similar to some previous studies, almost all their He-rich hot subdwarfs were found to be non-RV variable. Moreover, they discovered that He-poor hot subdwarfs in different regions of the $T_{\rm eff}$ -- $\log g$ diagram exhibit different RV-variability fractions. Cooler sdB stars (< 24,000 K) show a lower RV-variability fraction (${\rm 20^{+6}_{-4}}$\%) than sdB stars with $T_{\rm eff}$ $\sim$ 25,000 -- 32,000 K (${\rm 34^{+3}_{-3}}$\%). Hot subdwarfs with $T_{\rm eff}$ $\sim$ 32,000 -- 40,000 K (${\rm 34^{+3}_{-3}}$\%) and surface gravities ($\log g$) $\sim$ 5.6 -- 6.0 also show a lower RV-variability fraction (${\rm 22^{+4}_{-3}}$\%), while the largest RV-variability fraction for hot subdwarfs below the EHB was discovered (${\rm 46^{+6}_{-6}}$\%). 
   
   The RV variability of hot subdwarfs in observations is mainly influenced by their binary properties in short-period systems \citep[e.g.,][]{Geier.etal.2022}, which can help us to study the evolution of different types of hot subdwarfs and constrain their formation channels. To date, only \citet{Geier.etal.2022} have used a larger number of samples that contained different subclasses of hot subdwarfs to estimate the RV-variability fractions of hot subdwarfs along the different regions of the EHB. They discovered a different RV-variability fraction and origin for different subclasses of hot subdwarfs. However, the average uncertainty of the RV measurements in the study of \citet{Geier.etal.2022} is 18 km/s, which is somewhat large, and they only contained single-lined hot subdwarfs. To update the RV-variability fraction measurements of different subclasses of hot subdwarfs in \citet{Geier.etal.2022}, and to discuss their possible formation channels, we simultaneously use the multi-epoch spectra from the LAMOST DR11 medium-resolution spectroscopic (MRS) surveys and the low-resolution spectroscopic (LRS) surveys here to measure the RV variations of hot subdwarfs. The structure of this paper is as follows. In Sect. \ref{sect.2} we introduce the sample selection and the RV measurement method. Sect. \ref{sect.3} briefly describes our results. A comparison and discussion are presented in Sect. \ref{sect.4}, and our conclusions are given in Sect. \ref{sect.5}.
\section{Sample selection and radial velocity measurements}\label{sect.2}
\subsection{Sample selection}

   LAMOST is a 4m special reflecting Schmidt telescope with 4000 fibers. It is located at the Xinglong station of the National Astronomical Observatory \citep{Cui.etal.2012,Zhao.etal.2012}. Recently, LAMOST completed its regular survey observations for the first 11 years and published its 11th data release (DR11), which is not publicly available to all users. The LAMOST DR11 dataset contains 11,939,296 LRS and 13,187,912 MRS spectra. LAMOST LRS and MRS spectra have different spectral resolutions and wavelength ranges. LAMOST LRS spectra cover the wavelength range of 3690 -- 9100 $\AA$ with a resolution of 1800 at 5500 $\AA$, while LAMOST MRS spectra consist of two parts. The first part is the blue-arm spectrum, which covers 4950 -- 5350 $\AA$ with a resolution of 7500 at 5163 $\AA$, and the second part is the red-arm spectrum, which covers 6300 -- 6800 $\AA$ with a resolution of 7500 at 6593 $\AA$.

   More than 6,500 hot subdwarfs have so far been confirmed by spectroscopy \citep{Culpan.etal.2022,Geier.etal.2020}. Among them, many hot subdwarfs were identified through LAMOST LRS spectra \citep{Lei.etal.2020,Lei.etal.2019,Lei.etal.2018,Luo.etal.2021,Luo.etal.2020,Luo.etal.2019,Luo.etal.2016}, and these stars all have reliable atmospheric parameters, such as effective temperature, surface gravity, and surface helium abundance ($\log n{\rm (He)}/n{\rm (H)}$). In a recent study, \citet{Lei.etal.2023} newly identified 91 single-lined and 131 composite hot subdwarfs, which greatly increases the number of composite hot subdwarfs. All the hot subdwarfs identified in \citet{Lei.etal.2023,Lei.etal.2020,Lei.etal.2019,Lei.etal.2018} and \citet{Luo.etal.2021,Luo.etal.2020,Luo.etal.2019,Luo.etal.2016} were adopted as our initial samples. First, we cross-matched all the samples with the LAMOST DR11 MRS dataset. Hot subdwarfs with two or more observations in different nights were selected, and the criterion of the spectral signal-to-noise ratio (S/N) higher than 5 was employed to remove the poor-quality MRS spectra. Second, we cross-matched the remaining objects with the LAMOST DR11 LRS dataset. The spectral selection criteria for LRS are similar to those for MRS spectra, but with an S/N > 10 in the $g$ band for LRS spectra.

\subsection{Classifying the sample}

\subsubsection{Classification through the spectral energy distribution}

   Our sample consists of single-lined and composite hot subdwarfs. These two types of objects have distinct characteristics. Composite hot subdwarfs can show double-lined spectroscopic features, but they are sometimes very weak or cannot be detected. It is more efficient to detect IR excess in their spectral energy distributions (SEDs) to distinguish these objects. Recently, by using the Virtual Observatory SED Analyzer\footnote{\url{http://svo2.cab.inta-csic.es/svo/theory/vosa/}} \citep[VOSA,][]{Bayo.etal.2008} of the Spanish Virtual Observatory (SVO), \citet{Solano.etal.2022} built the largest SED database\footnote{\url{http://svocats.cab.inta-csic.es/hsa2/}} of hot subdwarfs so far. It contains 3186 hot subdwarfs from the catalog of hot subdwarfs compiled by \citet{Geier.etal.2020}. Through the IR-excess characteristic of composite hot subdwarfs, \citet{Solano.etal.2022} classified 2469 stars as single-SED and 615 stars as composite-SED. To classify the single-lined and composite hot subdwarfs in our sample, we cross-matched our stars with the database of \citet{Solano.etal.2022}, and obtained 109 composite and 365 single-lined hot subdwarfs. For a small portion of our hot subdwarfs that were not included in the database of \citet{Solano.etal.2022}, we used VOSA to build their SEDs and classified them. The observed photometric data were collected from the following catalogs: the Galaxy Evolution Explorer \citep[GALEX,][]{Martin.etal.2005}, the Panoramic Survey Telescope and Rapid Response System \citep[Pan-STARRS DR2,][]{Magnier.etal.2020}, Gaia \citep{Gaia.2020}, SDSS DR12 \citep{Alam.etal.2015}, the Two Micron All Sky Survey \citep[2MASS,][]{Skrutskie.etal.2006}, and the Wide-field Infrared Survey Explorer \citep[WISE,][]{Wright.etal.2010}. 
   
\subsubsection{Spectral classification}

   According to the traditional spectral classification scheme, hot subdwarfs can be divided into sdB, sdOB, sdO, He-sdB, He-sdOB, and He-sdO stars \citep{Moehler.etal.1990,Geier.etal.2017}. The classification criteria are as follows. SdB, sdOB, and sdO stars all have prominent hydrogen Balmer lines. SdB stars also show either no or weak He I lines, and sdOB stars usually show both weak He I and He II lines. SdO stars also have stronger He II lines than the He I lines, but sometimes, no He I lines can be detected. On the other hand, all He-sdB and He-sdOB stars exhibit prominent He I lines. Additionally, He-sdOB stars have weak He II and hydrogen Balmer lines, while no He II lines can be detected in He-sdB stars. He-sdO stars show prominent He II lines together with weak or no He I and hydrogen Balmer lines.
   
   Most of our hot subdwarfs were classified in the studies by \citet{Lei.etal.2023,Lei.etal.2020,Lei.etal.2019,Lei.etal.2018} and \cite{Luo.etal.2016} based on the spectral classification scheme described above. We adopted the classifications of these stars from the previous research. Additionally, a few hot subdwarfs from \citet{Luo.etal.2019,Luo.etal.2021} were not classified through the spectral line features. We inspected their spectra and classified them according to the traditional scheme.

\subsection{Radial velocity measurements}

   Binary stars often show RV variations due to the Doppler effect that can be measured through spectral lines shifts. The cross-correlation function (CCF) method is often used to measure the RVs. This method calculates the RVs by cross-correlating template spectra with observed spectra \citep{Huahui.etal.2023,Lichunqian.etal.2021,Sana.etal.2013,Merle.etal.2017,ZhangBo.etal.2021}. We used the algorithm of the CCF method as improved by \citet{ZhangBo.etal.2020,ZhangBo.etal.2021} to calculate the RVs. Since the sharp H$\alpha$ lines and He I lines usually appear in the red arm of LAMOST MRS spectra and sometimes no obvious lines can be found in the blue-arm spectra, the red-arm MRS and LRS spectra were adopted as observed spectra. 

        \begin{figure}[htbp]
    {
    \centering
    \includegraphics[width=0.49\textwidth]{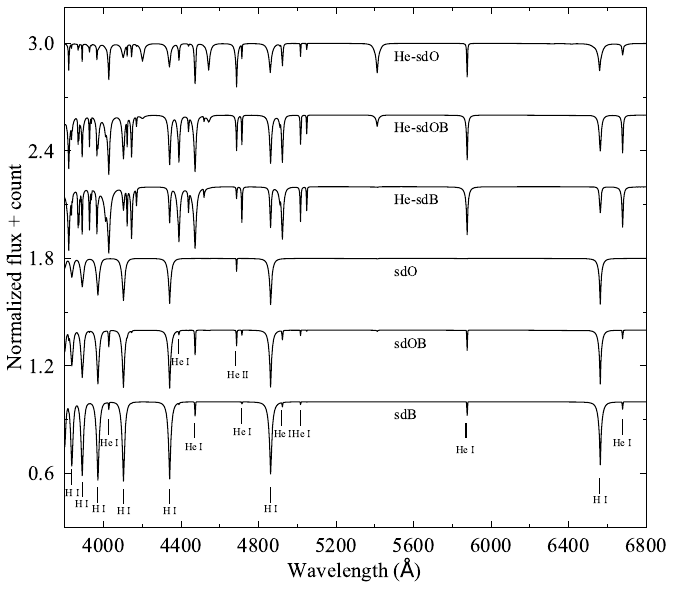}
      \caption{Selected template spectra for six subtypes of hot subdwarfs from their synthetic spectra.  
              }
         \label{Fig.1}
         }
    \end{figure}

    \begin{figure*}[htbp]
    {
    \centering
    \includegraphics[width=0.44\textwidth]{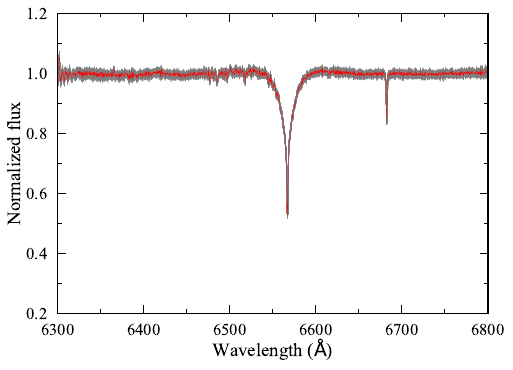}
    \includegraphics[width=0.43\textwidth]{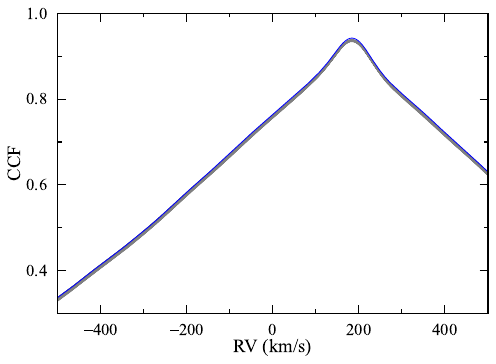}
      \caption{Spectra normalization of the observed and simulated spectra together with their corresponding CCF curves. Left panel: Normalized observed spectrum (red line) and the 100 normalized simulated spectra (gray lines). Right panel: Original CCF curve (blue line) and the CCF curves of the 100 simulated spectra (gray lines).
              }
         \label{Fig.2}
         }
    \end{figure*}  
   
   The synthetic spectra \citep[SYNSPEC version 49,][]{Lanz.etal.2007} of hot subdwarfs calculated from nonlocal thermodynamic equilibrium (NLTE) TLUSTY model atmospheres \citep[version 204,][]{Hubeny.etal.2017} were adopted as our template. The template and observed spectra were normalized through the package LASPEC \citep{ZhangBo.etal.2021}. Based on the spectral classification, our hot subdwarfs were divided into six subtypes. In order to accurately measure their RVs, we selected one corresponding template for each type of hot subdwarf from their synthetic spectra. The classification of template spectra also followed the traditional scheme \citep{Moehler.etal.1990,Geier.etal.2017}. Our selected template spectra are shown in Fig. \ref{Fig.1}. Except for the hydrogen Balmer lines, the template spectra of sdB, sdOB, and sdO stars also have He I lines, both He I and He II lines, and He II lines, respectively. Except for the prominent He I lines for He-sdB and He-sdOB stars, and the prominent He II lines for He-sdO stars, the template spectra of He-sdOB and He-sdO stars also have He II and He I lines, respectively. Moreover, the template spectra of He-sdB, He-sdOB, and He-sdO stars all present detectable hydrogen Balmer lines. We cross-correlated the observed spectra of each type of hot subdwarf with their corresponding template, and the range of RV variations was set from $-500$ to 500 km/s with an interval of 1 km/s.

   The spectral classification of our composite hot subdwarfs and the processes for measuring their RVs are the same as those that were used for single-lined objects. Some of our composite hot subdwarfs present double-lined features, but others only show an IR excess in their SEDs, and no spectral lines from the companions can be detected. In addition, the luminous hot subdwarfs contribute the majority of the flux for composite systems. Many double-lined composite hot subdwarfs show only weak lines of F-, G-, or K-type companions in the LAMOST LRS spectra. In this paper, we mainly focus on the primary component, and the RVs were only derived from the primary.  

\subsection{Radial velocity uncertainties} 

   The Monte Carlo method was employed to estimate the uncertainties of our RV measurements \citep{Huahui.etal.2023,Lichunqian.etal.2021}. For each flux at each wavelength point of the observed spectrum, 100 simulated flux values were randomly generated from a Gaussian distribution based on the flux and flux error. This allowed us to generate 100 simulated spectra for each observed spectrum. The 100 simulated spectra were used to calculate their CCF curves and RVs, and the standard deviation of the 100 RVs was taken as the RV uncertainty. The left panel of Fig. \ref{Fig.2} shows the normalization of an observed red-arm MRS spectrum and 100 simulated spectra. Its right panel presents the original CCF curve and the 100 CCF curves derived through the simulated spectra. Since the systematic uncertainties are also induced by the stability of LRS/MRS over time spans of years, the spectral processing procedure, and so on, we added a systematic uncertainty of 5 km/s in quadrature to all the RV uncertainties measured from the LRS/MRS spectra to calculate their final uncertainties \citep{Geier.etal.2024,Geier.etal.2022}. 
    
\begin{figure}[htbp]
    {
    \centering
    \includegraphics[width=0.46\textwidth]{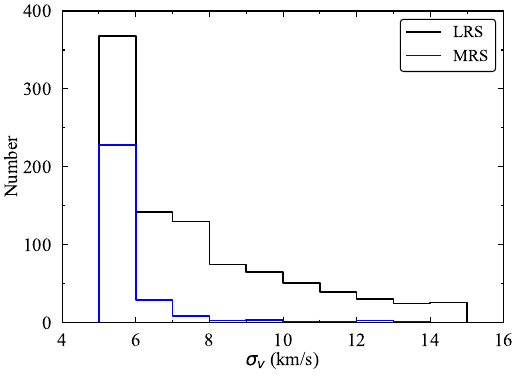}
      \caption{Radial velocity uncertainty distributions of hot subdwarfs. The black line shows the RV uncertainties measured from the LRS spectra, and the blue line represents the RV uncertainties measured from the MRS spectra.
              }
         \label{Fig.3}
         }
    \end{figure}

   To obtain as many reliable RVs as possible and to explore the RV-variability fractions of different subclasses of hot subdwarfs with relatively smaller RV uncertainties, we set a limit to select stars with RV uncertainties smaller than 15 km/s. Furthermore, the CCF maximum of each observed spectrum was restricted to be larger than 0.15 \citep{Merle.etal.2017}. Eventually, we obtained the multi-epoch RVs for 434 hot subdwarfs, of which 45 are from the MRS spectra and 389 are from the LRS spectra. Fig. \ref{Fig.3} presents the distributions of $\sigma_{\textup{v}}$ for these stars. The average $\sigma_{\textup{v}}$ obtained from the LRS and MRS spectra is 7.6 km/s and 5.6 km/s, respectively, and the average $\sigma_{\textup{v}}$ obtained from all the LRS and MRS spectra is 7.1 km/s.

\section{Distribution of radial velocity variations and variability criterion}\label{sect.3}
\subsection{Distribution of radial velocity variations}\label{sect.3.1}

    \begin{figure*}[htbp]
    {
    \centering
    \includegraphics[width=0.48\textwidth]{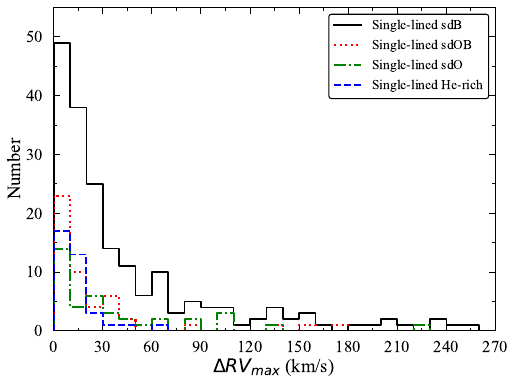}
    \includegraphics[width=0.48\textwidth]{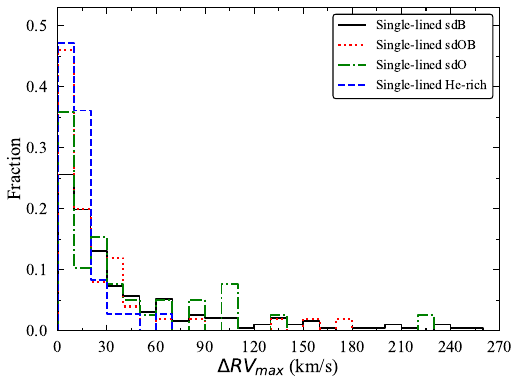}
    \includegraphics[width=0.48\textwidth]{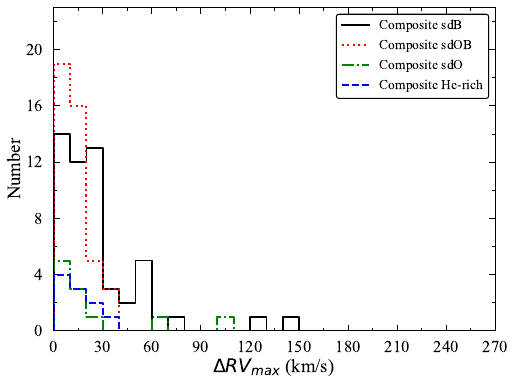}
    \includegraphics[width=0.48\textwidth]{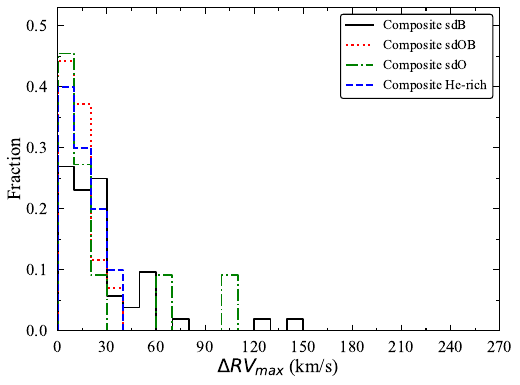}
      \caption{$\Delta RV_{\textup{max}}$ distributions of different types of hot subdwarfs. Left panels: Absolute numbers in each bin. Right panels: Normalized fractions in the same bin as in the left panels. The solid black, dotted red, dash-dotted green, and dashed blue lines in the upper panels represent the $\Delta RV_{\textup{max}}$ distributions for single-lined sdB, sdOB, sdO, and He-rich hot subdwarfs. The same types of lines in the bottom panels show the $\Delta RV_{\textup{max}}$ distributions of composite sdB, sdOB, sdO, and He-rich hot subdwarfs, respectively.
              }
         \label{Fig.4}
         }
    \end{figure*}

    \begin{table*}[htp]
    \fontsize{9.5pt}{11pt}\selectfont
    \caption[]{RV-variability fractions of different types of hot sundwarfs and a comparison with \citet{Geier.etal.2022}.}
    
    \label{Table.1}
    \begin{tabular}{ccccccccc}
    \hline
    \noalign{\smallskip}
    Subtype  & Total & RV-variable & Fraction (\%) & Total (Geier) & RV-variable (Geier) & Fraction (Geier) (\%) \\ \hline
    \noalign{\smallskip}
    all single-lined   & 317 & 79 & 25 $\pm$ 2 & 646 & 164 & 25 $\pm$ 2  \\
    sdB      & 191 & 60  & 31 $\pm$ 3 & 378 & 120 & 32 $\pm$ 2 \\
    sdOB     & 51  & 7  & 14 $\pm$ 5 & 117 & 29 & 25 $\pm$ 4 \\
    sdO      & 39  & 10  & 26 $\pm$ 7 &  44 & 12 & 27 $\pm$ 7 \\
    He-rich   & 36 & 2 & 6 $\pm$ 4 & 107 & 3 & 3 $\pm$ 2 \\
    \hline
    \noalign{\smallskip}
    all composite  & 117 & 10 & 9 $\pm$ 3 & -- & -- & -- \\
    sdB      & 52  & 8 & 15 $\pm$ 5 & -- & -- & -- \\
    sdOB   & 44  & 0 & $0^{+1}_{-0}$  & -- & -- & -- \\ 
    sdO    & 11  & 2 & 18 $\pm$ 12 & -- & -- & -- \\  
    He-rich    & 10  & 0 & $0^{+5}_{-0}$ & -- & -- & -- \\  \hline
    \noalign{\smallskip}    
    \end{tabular}
    \end{table*} 

   Our final sample consisted of 317 single-lined and 117 composite hot subdwarfs. The maximum RV variation ($\Delta RV_{\textup{max}}$) was calculated by subtracting the minimum from the maximum RV of the star. In the upper panels of Fig. \ref{Fig.4}, we show the $\Delta RV_{\textup{max}}$ distributions of single-lined sdB, sdOB, sdO, and He-rich hot subdwarfs. All single-lined He-rich hot subdwarfs have helium abundances of $\log n{\rm (He)}/n{\rm (H)} > -1$. We found that single-lined He-rich hot subdwarfs obviously exhibit more fractions of stars with $\Delta RV_{\textup{max}}$ lower than 20 km/s compared to single-lined sdB stars. In order to assess the statistical difference of this comparison, the Kolmogorov-Smirnov (K-S) test was employed. We obtained a $ P_{\textup{KS}}$ value of $9\times 10^{-6}$, which indicates that the $\Delta RV_{\textup{max}}$ distributions of single-lined sdB and He-rich hot subdwarfs are not from the same mother sample. Moreover, more than 600 He-rich hot subdwarfs were discovered, but only a few special He-rich hot subdwarfs were found in short-period binaries \citep{Ahmad.etal.2004,Sener.etal.2014,Lisker.etal.2004,Ratzloff.etal.2020,Reindl.etal.2020,Kupfer.etal.2017a,Kupfer.etal.2020a,Kupfer.etal.2020b,Snowdon.etal.2023}. Most of them are very special stars, which may originate from the CE ejection of intermediate-mass stars when they cross the Hertzsprung gap \citep{Kupfer.etal.2020a,Kupfer.etal.2020b} or come from the CE ejection of post-early asymptotic giant branch stars \citep[AGB,][]{Li.etal.2024}. The majority of He-rich hot subdwarfs are very unlikely to form through these rare and peculiar channels. Our results support the opinions of \citet{Geier.etal.2022} and other researchers that most He-rich hot subdwarfs may be single stars, and merger channels such as the double He-WDs merger may lead to their formation, which can explain their surface helium enrichment and $\Delta RV_{\textup{max}}$ distributions well.
   
   In the bottom panels of Fig. \ref{Fig.4}, we show the $\Delta RV_{\textup{max}}$ distributions of composite sdB, sdOB, sdO, and He-rich hot subdwarfs. We found that composite sdB and sdOB stars exhibit a smaller fraction of stars with $\Delta RV_{\textup{max}}$ higher than 60 km/s than their corresponding type of single-lined hot subdwarfs. The KS-test showed a statistically significant difference in the $\Delta RV_{\textup{max}}$ distributions between composite sdB and single-lined sdB stars ($ P_{\textup{KS}}$  = $4.9\times 10^{-4}$), which indicates that these two types of stars belong to different populations. Nevertheless, except that no composite sdOB stars show $\Delta RV_{\textup{max}}$ higher than 60 km/s, the $\Delta RV_{\textup{max}}$ distributions in 0 -- 40 km/s for composite sdOB and single-lined sdOB stars are similar. The KS-test does not reveal a statistically significant difference in the $\Delta RV_{\textup{max}}$ distributions between these two types of stars either ($ P_{\textup{KS}}$  = 0.20). Based on the limited RV accuracy achieved by LAMOST LRS/MRS spectra, it is hard to detect an obvious difference between the $\Delta RV_{\textup{max}}$ distributions of composite and single-lined sdOB stars. More observations with higher-accuracy RVs may be needed to reveal whether their $\Delta RV_{\textup{max}}$ distributions differ. Because the sample size of composite sdO and composite He-rich hot subdwarfs is limited, the statistical analysis for these stars may not be reliable. We just present their $\Delta RV_{\textup{max}}$ distributions as a reference.
    

\subsection{Criterion for radial velocity variability}

   We adopted the same method as used by \citet{Maxted.etal.2001}, \citet{Geier.etal.2022,Geier.etal.2024}, \citet{Napiwotzki.etal.2020}, and \citet{Huahui.etal.2024} to estimate the probability of a star that is variable in RV. First, we calculated the weighted mean RV of every star and assumed the value as a constant. Then, we calculated the $\chi^{2}$ statistic. By comparing the calculated $\chi^{2}$ with the $\chi^{2}$ distribution for the appropriate degrees of freedom, we calculated the probability ($p$) of obtaining the observed $\chi^{2}$ value or a higher value from random fluctuations of a constant velocity. Stars that showed a false detection $p$ lower than 0.01\% ($\log p < -4$) were considered to be significantly RV-variable objects. In Table \ref{Table.A1} we present the RV variability and other parameters of our sample.

\subsection{Variability fractions in different types of hot subdwarfs}

    In the upper part of Table \ref{Table.1}, we list the RV-variability fractions of different types of single-lined hot subdwarfs based on our study and \citet{Geier.etal.2022}. The data catalog\footnote{\url{http://cdsarc.u-strasbg.fr/viz-bin/cat/J/A+A/661/A113}} of \citet{Geier.etal.2022} also contains the spectral classifications of 646 hot subdwarfs based on the scheme of \citet{Moehler.etal.1990}, which were adopted by us for the sample of \citet{Geier.etal.2022}, and hot subdwarfs with $\log n{\rm (He)}/n{\rm (H)} > -1$ were classified as He-rich. The uncertainties in the RV-variability fractions were derived based on the assumption of a binomial distribution, while the uncertainties for the subtypes without detected RV-variable stars were estimated by using the beta distribution to calculate the 68\% confidence upper limit. It should be noted that the RV-variability fractions of different types of hot subdwarfs are just their lower limits since some sources only have a small number of RV measurements, and the long-period hot subdwarf binaries are difficult to detect with significant RV variations. 
    
    The detected RV-variability fractions of our single-lined sdB, sdO, and He-rich hot subdwarfs are all consistent with the fractions from the sample of \citet{Geier.etal.2022} within the uncertainties. Only two of our single-lined He-rich hot subdwarfs are found to be RV variable. They are PG 1415+492 and BD+48 1777, and they exhibit RV variations of 64 km/s and 33 km/s, respectively. These two special stars may be in short-period binaries and currently lack available orbital solutions. They deserve further observations to study their formation. The RV-variability fraction of our single-lined sdOB stars is 14 $\pm$ 5\%, which is lower than the fraction of single-lined sdB stars. A somewhat higher RV-variability fraction of 25 $\pm$ 4\% is found in the single-lined sdOB stars of \citet{Geier.etal.2022}. It is slightly lower than that fraction of their single-lined sdB stars within the uncertainties. A detailed discussion of the RV variability of single-lined sdOB stars is presented in Sect. \ref{sect.4.1}. 
    
    The bottom part of Table \ref{Table.1} presents the RV-variability fractions of different types of composite hot subdwarfs. We found that 9 $\pm$ 3\% of all our composite hot subdwarfs exhibit significant RV variations, which is lower than the fraction of 25 $\pm$ 2\% for all single-lined hot subdwarfs. The RV-variability fractions of our composite sdB and sdOB are also lower than their corresponding type of single-lined stars. Currently, 27 composite hot subdwarfs possess solved orbital parameters \citep{Vos.etal.2012,Vos.etal.2013,Vos.etal.2017,Vos.etal.2019,Deca.etal.2012,Barlow.etal.2012,Barlow.etal.2013,Ostensen.etal.2012,Molina.etal.2022,Nemeth.etal.2021,Dorsch.etal.2021}. All of them are long-period binaries with periods longer than 400 days. Since the average uncertainty of our RV measurements from LAMOST spectra is about 7 km/s, the lower detected RV-variability fraction for composite hot subdwarfs is expected because most of them may be long-period binaries, which is consistent with the supposition of \citet{Copperwheat.etal.2011}. 

         \begin{figure*}[htbp]
    {
    \centering
    \includegraphics[width=0.486\textwidth]{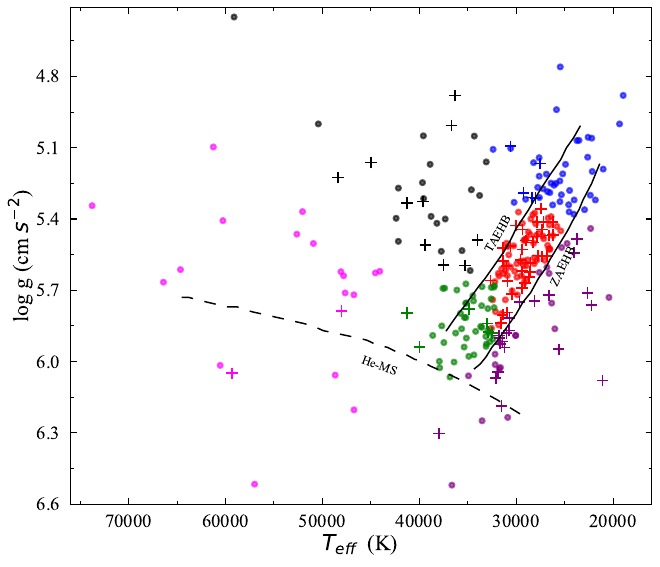}
    \includegraphics[width=0.484\textwidth]{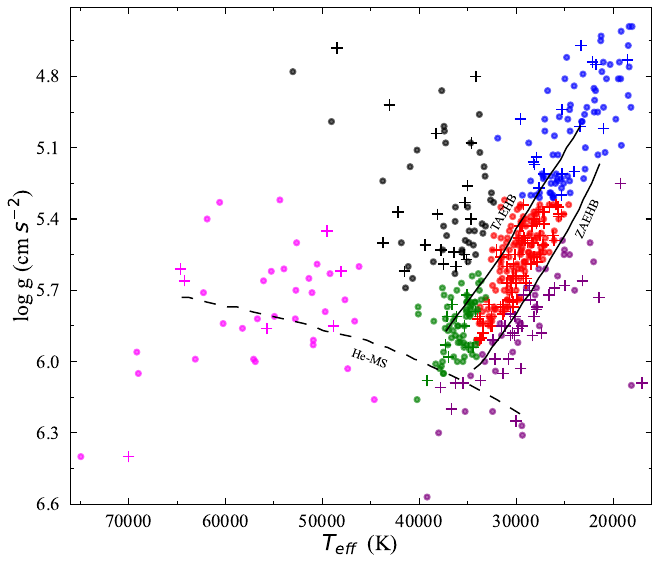}
      \caption{$T_{\rm eff}$ -- $\log g$ diagram of different subclasses of single-lined He-poor hot subdwarfs. The left panel shows our sample, and the right panel shows the sample of \citet{Geier.etal.2022}. The circles represent non-RV-variable hot subdwarfs, and the crosses represent RV-variable hot subdwarfs. The sample is divided into six subclasses that are marked by different colors, for which EHBa, EHBb, EHBc, bEHB, postEHBa, and postEHBb are plotted in blue, red, green, purple, black, and magenta, respectively. The zero-age EHB (ZAEHB) and terminal-age EHB (TAEHB) sequences with Z = 0.02 from \citet{Dorman.etal.1993} are labeled by solid black lines. The helium main sequence (He-MS) from \citet{Paczynski.etal.1971} is marked by the dashed black line.
              }
         \label{Fig.5}
         }
    \end{figure*} 

    \begin{table*}[htp]
    \fontsize{9.5pt}{11pt}\selectfont
    \caption[]{RV-variability fractions for the six subclasses of hot subdwarfs and a comparison with \citet{Geier.etal.2022}.
    }
    \label{Table.2}
    \begin{tabular}{cccccccc}
    \hline
    \noalign{\smallskip}
    Subclass      & Total &   RV-variable   &   Fraction (\%)  & Total(Geier) &   RV-variable (Geier)  &   Fraction (Geier) (\%)  \\ \hline
    \noalign{\smallskip}
    EHBa      & 47 & 5  & 11 $\pm$ 4 & 92 & 16  & 17 $\pm$ 4  \\
    EHBb    & 96  & 33  & 34 $\pm$ 5 & 195 & 72  & 37 $\pm$ 3  \\
    EHBc    & 48  & 6 & 13 $\pm$ 5 & 84 & 11  & 13 $\pm$ 4 \\ 
    bEHBa    & 41  & 21 & 51 $\pm$ 8 & 69 & 36  & 52 $\pm$ 6 \\ 
    postEHBa    & 29  & 10 & 34 $\pm$ 9 & 60 & 19  & 32 $\pm$ 6 \\
    postEHBb    & 20  & 2 & 10 $\pm$ 7 & 39 & 7  & 18 $\pm$ 6 \\ \hline
    \noalign{\smallskip}    
    \end{tabular}
    \end{table*}

    Our results support that most composite hot subdwarfs are in long-period binaries, and the stable RLOF channel may be their main formation channel \citep[e.g.,][]{Pelisoli.etal.2020,Vos.etal.2020}. However, there still exist some special cases, in which 8 of our composite sdB stars and 2 of our composite sdO stars exhibit significant RV variations. These stars all show RV variations larger than 45 km/s, and LAMOSTJ 073756.25+311646.5, PG 1629+081, and GALEXJ 071314.5+173459 all show RV variations larger than 100 km/s. Similar to our study, 5 composite sdB stars were also detected with significant RV variations in the 32 composite sdB stars (16\%) of \citet{Copperwheat.etal.2011}. The RV variations of these composite hot subdwarfs might be caused by the following several reasons. First, the evolutionary pathways from triple systems were proposed to produce hot subdwarfs, and a great number of hot subdwarfs from these scenarios were predicted to be in triple systems \citep{Preece.etal.2022}. In observations, several triple-system candidates were also discovered \citep{Heber.etal.2002,Kupfer.etal.2015,Pelisoli.etal.2020,Schaffenroth.etal.2023}. These stars are composed of an inner short-period hot subdwarf binary with an unseen companion and a distant F-, G-, or K-type companion. The RV-variable composite hot subdwarfs could be in such triple systems, and they might be triple-system candidates. Second, some of these stars might be blends, and they are not real composite hot subdwarf binaries. Third, although all the composite hot subdwarfs with solved orbital parameters were proven to be long-period binaries, we still cannot exclude the possibility that these RV-variable composite hot subdwarfs are in short-period binaries with F-, G-, or K-type companions \citep{Copperwheat.etal.2011}. Because we only have a few RV measurements of these composite hot subdwarfs, the nature of the companions cannot be defined. More follow-up observations are needed to reveal the properties of these stars.

\section{Comparison and discussion}\label{sect.4}
\subsection{Distribution in the $T_{\rm eff}$ -- $\log g$ diagram}\label{sect.4.1}
    
   \citet{Geier.etal.2022} discovered that the RV-variability fractions of single-lined He-poor hot subdwarfs in different regions of the $T_{\rm eff}$ -- $\log g$ diagram are different. In Fig. \ref{Fig.5} we compared the distributions of our single-lined He-poor hot subdwarfs with those of \citet{Geier.etal.2022}. All the atmospheric parameters of our hot subdwarfs are from the previously identified works \citep{Lei.etal.2023,Lei.etal.2020,Lei.etal.2019,Lei.etal.2018,Luo.etal.2021,Luo.etal.2020,Luo.etal.2019,Luo.etal.2016}. Similar to the division criteria of \citet{Geier.etal.2022} but with some differences, we divided hot subdwarfs into six subclasses (EHBa, EHBb, EHBc, bEHBa, postEHBa, and postEHBb) through visual inspection of their locations in the $T_{\rm eff}$ -- $\log g$ diagram and the density of the RV-variable together with non-RV-variable stars, as well as their spectral types. A detailed discussion is given in the following subsections. The statistics for the RV-variability fractions of these subclasses from our sample and that of \citet{Geier.etal.2022} are shown in Table \ref{Table.2}. Our RV-variability fractions for the six subclasses are consistent with those of \citet{Geier.etal.2022} within the uncertainties, although we implemented a stricter restriction that limited the RV uncertainties to less than 15 km/s.
   
\subsubsection{Stars along the extreme horizontal branch}
   
   As shown in Fig. \ref{Fig.5}, EHBa, EHBb, and EHBc stars are all distributed along the EHB. Almost all the EHBa and EHBb are sdB stars. There seems to be a change in the density of RV-variable stars at $T_{\rm eff}$ $\sim$ 25,000 K and $\log g$ $\sim$ 5.3, which divides the majority of sdB stars into EHBa and EHBb. The RV-variability fraction of EHBa stars based on our sample and on that of \citet{Geier.etal.2022} is 11 $\pm$ 4\% and 17 $\pm$ 4\%. Both values are lower than the fraction of 34 $\pm$ 5\% and 37 $\pm$ 3\% for EHBb stars. \citet{Geier.etal.2022} first noted that the cooler sdB stars present a lower RV-variability fraction than the higher-temperature group. This distribution trend for RV-variable sdB stars is consistent with the prediction of BPS simulations from \citet{Han.etal.2003} that the CE ejection channel can only form a few sdB stars cooler than about 25,000 K. Based on the lower RV-variability fraction and the BPS simulations, \citet{Geier.etal.2022} speculated that the cooler group of sdB stars may contain a new subpopulation of long-period binaries with late-type MS or compact companions. Moreover, the simulations of the stable RLOF channel from \citet{Vos.etal.2020} indicated that about one-third of the sdB stars from this channel are long-period single-lined binaries with late-type MS companions, and sdB stars with $T_{\rm eff}$ $\sim$ 20,000 K can be produced. This further supports the supposition of \citet{Geier.etal.2022}. In addition, some of the cooler sdB stars may also have been created through merger channels, such as the merger of an RGB star with its companion \citep{Politano.etal.2008}.

\begin{figure}[htbp]
    {
    \centering
    \includegraphics[width=0.46\textwidth]{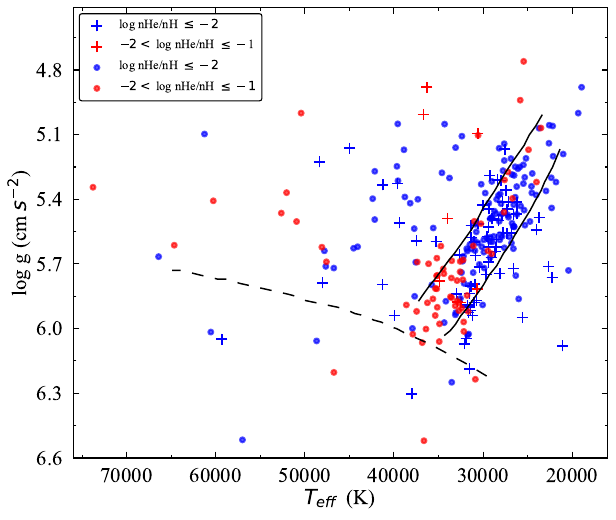}
      \caption{Distributions of single-lined He-poor hot subdwarfs with different helium abundances. The hot subdwarfs with helium abundances of $\log n{\rm (He)}/n{\rm (H)}$ in the two different ranges are labeled by different colors. Same as Fig. \ref{Fig.5}, the circles represent non-RV-variable hot subdwarfs, and the crosses represent RV-variable hot subdwarfs.
              }
         \label{Fig.6}
         }
    \end{figure}

   The classification for EHBb and EHBc stars is based on their spectral types. A boundary exists between sdB and sdOB stars at $T_{\rm eff}$ $\sim$ 33,000 K. Hot subdwarfs in the region of the sdOB stars grouping are classified as EHBc, and they also contain a few sdB stars. We separated the EHBc and postEHBa stars through the small density gap of sdOB stars at $\log g$ $\sim$ 5.7. The RV-variability fraction of EHBc stars (with $T_{\rm eff}$ $\sim$ 33,000 -- 40,000 K and $\log g$ $\sim$ 5.7 -- 6.0) from our sample and that of \citet{Geier.etal.2022} shows a lower value of 13 $\pm$ 5\% and 13 $\pm$ 4\% compared to EHBb stars. \citet{Geier.etal.2022} also found that hot subdwarfs of their EHB3 group, which are located in a similar region to our EHBc stars, exhibited a lower RV-variability fraction than their EHB2 group. They suggested two possible reasons for the lower RV-variability fraction. The first reason was that their EHB3 sample might be populated by some single iHe-sdOB stars with inaccurate helium abundance measurements. However, as shown in Fig. 3 of \citet{Luo.etal.2021}, they obtained quite consistent helium abundance measurements with \citet{Lei.etal.2018,Lei.etal.2019,Lei.etal.2020} ranging from $-2 < \log n{\rm (He)}/n{\rm (H)} < 0$. In Fig. \ref{Fig.6} we present the distributions of our single-lined hot subdwarfs in the two different helium abundance ranges. The figure clearly shows that most EHBc stars have helium abundances in the range of  $-2 < \log n{\rm (He)}/n{\rm (H)} \leq -1$. We therefore think that the majority of helium abundance measurements for our EHBc stars are accurate enough. The alternative explanation provided by \citet{Geier.etal.2022} is that the non-RV-variable stars in their EHB3 group might evolve from single iHe-rich hot subdwarfs by diffusion processes as proposed by \citet{Miller.etal.2008}. Most of our EHBc stars exhibit higher helium abundances of $\log n{\rm (He)}/n{\rm (H)}$ than EHBb stars. This is consistent with the prediction of \citet{Miller.etal.2008} based on diffusion processes. The higher helium abundances as well as the lower RV-variability fraction of EHBc stars all support an evolutionary connection between iHe-rich hot subdwarfs and EHBc stars. 

    \begin{figure*}[htbp]
    \centering
   \begin{minipage}{0.32\textwidth}
   \includegraphics[width=\textwidth]{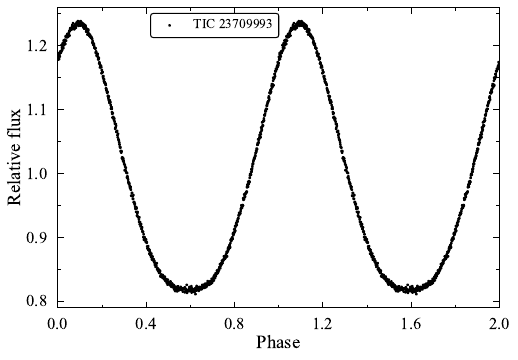}
   \end{minipage}
    \begin{minipage}{0.32\textwidth}
    \includegraphics[width=\textwidth]{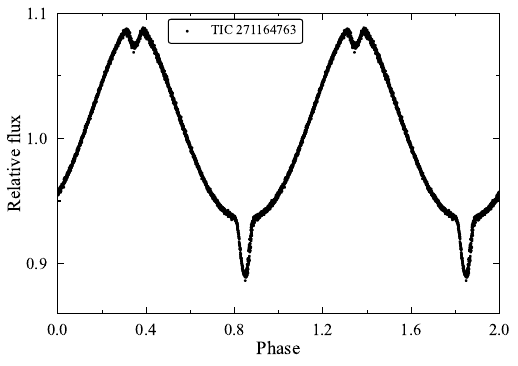}
    \end{minipage}
    \begin{minipage}{0.32\textwidth}
    \includegraphics[width=\textwidth]{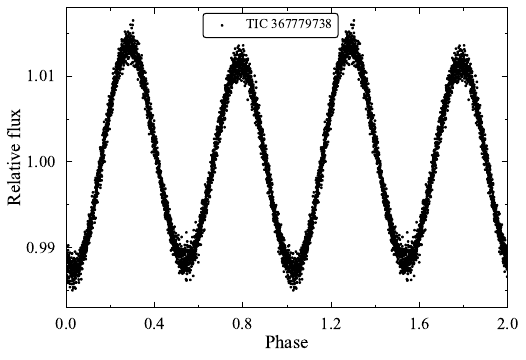}
    \end{minipage}
   \caption{Light curves of different types of light-variable hot subdwarfs. The light curves of a reflection effect and a HW Vir system in the first and second panels are phase-folded to the orbital period. The third panel shows the light curve of an ellipsoidal deformation system phase-folded to twice the orbital period.}
    \label{Fig.7}
    \end{figure*}
    
       \begin{figure*}[htbp]
    {
    \centering
    \includegraphics[width=0.46\textwidth]{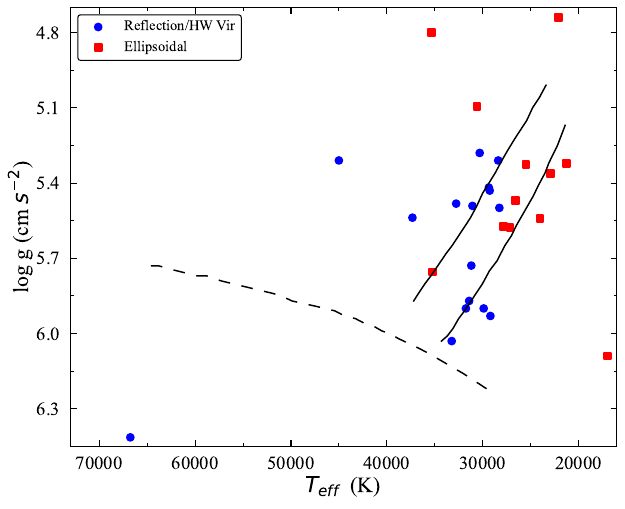}
    \includegraphics[width=0.46\textwidth]{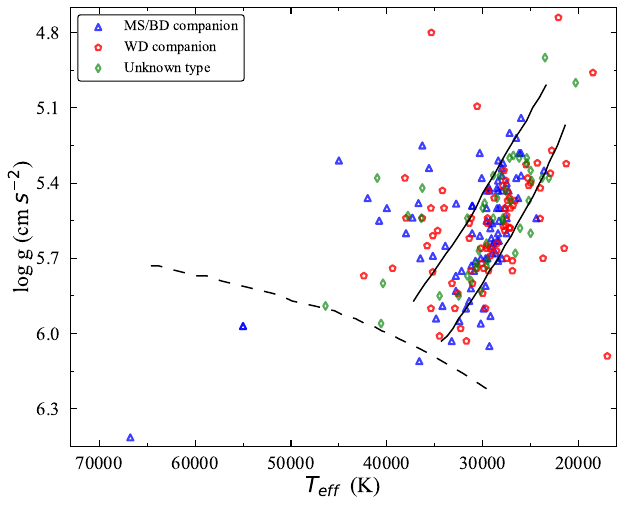}
    \caption{$T_{\rm eff}$ -- $\log g$ diagram of the confirmed short-period binary hot subdwarfs from light curves and the literature. Left panel: Newly found reflection effect or HW Vir systems (blue circles) and the newly discovered ellipsoidal deformation systems (red squares). Right panel: Known short-period binary hot subdwarfs in the literature and our newly found short-period binaries through light curves. Hot subdwarfs with MS or BD companions and WD companions are labeled as blue triangles and red pentagons, respectively. Short-period binary hot subdwarfs with an unknown companion type are labeled as green diamonds.}
    \label{Fig.8}
    }
    \end{figure*}

\subsubsection{Stars above and beyond the extreme horizontal branch}

   The division between postEHBa and EHBa, b, and c stars is mainly based on their density and locations. Hot subdwarfs located above the EHB and separate from EHBa, b, and c stars are classified as postEHBa. There seems to be a density gap of stars at $T_{\rm eff}$ $\sim$ 45,000 K that divides hot subdwarfs above and beyond the EHB into two subclasses. The lower-temperature group with $T_{\rm eff}$ $\sim$ 35,000 -- 45,000 K is labeled as postEHBa, and the higher-temperature group with $T_{\rm eff}$ $\sim$ 45,000 -- 70,000 K is labeled as postEHBb. We found that the RV-variability fraction of postEHBa stars from our sample and that of \citet{Geier.etal.2022} is 34 $\pm$ 9\% and 34 $\pm$ 6\%. Both values are close to the fraction of EHBb stars. Since only a few postEHBa stars are predicted to form through the CE ejection channel based on the BPS simulations from \citet{Han.etal.2003}, these stars may have an evolutionary connection to EHBb stars. Nevertheless, postEHBb stars show a lower RV-variability fraction of 10 $\pm$ 7\% and 18 $\pm$ 6\% based on our sample and on that of  \citet{Geier.etal.2022}. The evolutionary process that begins from EHBb stars and then connects to postEHBb stars might not apply to all postEHBb stars. One possible scenario for their formation is that some non-RV-variable postEHBb stars might evolve from the single He-sdO stars by diffusion processes \citep{Miller.etal.2008}. 
   
\subsubsection{Stars below the extreme horizontal branch}

   Hot subdwarfs below the canonical ZAEHB are classified as bEHBa. Because the location of ZAEHB derived through the evolutionary tracks has some uncertainties that depend on the metallicity and helium-core mass of the stars \citep{Dorman.etal.1993,Geier.etal.2022}, we did not strictly constrain our bEHBa stars to be below the ZAEHB. Some stars on the line or very close to it were also classified as bEHBa. The bEHBa stars based on our sample and \citet{Geier.etal.2022} show the largest RV-variability fraction of 51 $\pm$ 8\% and 52 $\pm$ 6\%. \citet{Geier.etal.2022} also noted the high RV-variability fraction of hot subdwarfs below the EHB. They suggested that the concentration of hotter stars (> 25,000 K) below the EHB might be composed of low-mass hot subdwarfs that evolved from the intermediate-mass stars with ignited nondegenerate helium cores, while several more widely distributed cooler stars might be progenitors of He-WDs with non-core helium burning (a detailed discussion can be seen in \citet{Geier.etal.2022}). Based on the high RV-variability fraction and the locations of our bEHBa stars, we support that the majority of these stars might be low-mass hot subdwarfs, especially the concentration of bEHBa stars at $T_{\rm eff}$ $\sim$ 30,000 K, while the cooler bEHBa stars (< 25,000 K) and the bEHBa stars below the He-MS might be progenitors of He-WDs.

\subsection{Comparison with short-period binary hot subdwarfs from light curves and the literature}
\subsubsection{Collecting short-period binaries}
    
   Many hot subdwarfs show different types of variability in their light curves due to binarity or pulsations. Short-period binary hot subdwarfs with low-mass MS or BD companions can exhibit quasi-sinusoidal variability in their light curves owing to the reflection effect, and HW Vir systems are eclipsing reflection effect systems \citep{Schaffenroth.etal.2022,Schaffenroth.etal.2023,Kevin.etal.2023,Baran.etal.2021,Barlow.etal.2022}. Compact hot subdwarf binaries with WD companions can show ellipsoidal deformation \citep{Pelisoli.etal.2021,Lin.etal.2024,Kupfer.etal.2022}. Additionally, many hot subdwarfs show variability due to pulsations, with typical periods of a few minutes or about one to two hours \citep{Kupfer.etal.2021,Baran.etal.2023,Baran.etal.2024}. 
   
   Space-based telescope observations like the Transiting Exoplanet Survey Satellite \citep[TESS,][]{Ricker.etal.2015} and the K2 space missions \citep{Howell.etal.2014} are very efficient in obtaining a large number of high-quality light curves for many stars. \citet{Schaffenroth.etal.2022} used light curves from TESS and K2 to search for short-period binary hot subdwarfs. They detected 82 new sdB+dM/BD and 23 new sdB+WD systems. It is quite interesting to investigate whether there is any difference for the distributions in the $T_{\rm eff}$ -- $\log g$ diagram between RV-variable hot subdwarfs and other confirmed short-period binary hot subdwarfs. To achieve this comparison, we used the 2-minute cadence and 20-second cadence light curves from TESS to search for short-period binary hot subdwarfs in the single-lined samples of \citet{Lei.etal.2018,Lei.etal.2019,Lei.etal.2020,Lei.etal.2023,Luo.etal.2016,Luo.etal.2019,Luo.etal.2020,Luo.etal.2021}, and \citet{Geier.etal.2022}. Following the data reduction method used by \citet{Schaffenroth.etal.2022}, we adopted the publicly available custom script\footnote{\url{https://github.com/ipelisoli/TESS-LS}} developed by \citet{Pelisoli.etal.2020} to download light curves and generate periodograms. Subsequently, we phase-folded the light curves to the period determined by the periodogram or twice the period for ellipsoidal deformation systems. Example light curves of a reflection effect, a HW Vir, and an ellipsoidal deformation system are presented in Fig. \ref{Fig.7}. After conducting visual inspections for the light curves of 593 hot subdwarfs, we newly discovered 12 reflection effect, 4 HW Vir, and 12 ellipsoidal deformation systems that were previously unknown. These stars are plotted in the left panel of Fig. \ref{Fig.8}. In Table \ref{Table.A2} we present the period and other parameters of these stars. 
   
   The right panel of Fig. \ref{Fig.8} shows the distributions of known short-period binary hot subdwarfs with atmospheric parameters in the literature \citep[and references therein]{Schaffenroth.etal.2022,Schaffenroth.etal.2019,Schaffenroth.etal.2018,Schaffenroth.etal.2015,Schaffenroth.etal.2013,Baran.etal.2019,Kupfer.etal.2017b,Kupfer.etal.2020b,Ratzloff.etal.2019,Ostensen.etal.2010,Kawka.etal.2015,Geier.etal.2014,Geier.etal.2011,Geier.etal.2010,Copperwheat.etal.2011,Maxted.etal.2001,Kupfer.etal.2015}, as well as our newly discovered short-period binaries through light curves. The companions of short-period binary hot subdwarfs are mainly dM, BD, or WD stars. We also present the distributions of these two types of hot subdwarfs by blue triangles and red pentagons. We found that they exhibit a quite similar distribution in the $T_{\rm eff}$ -- $\log g$ diagram.

\subsubsection{Comparison}

\begin{figure}[htbp]
    {
    \centering
    \includegraphics[width=0.47\textwidth]{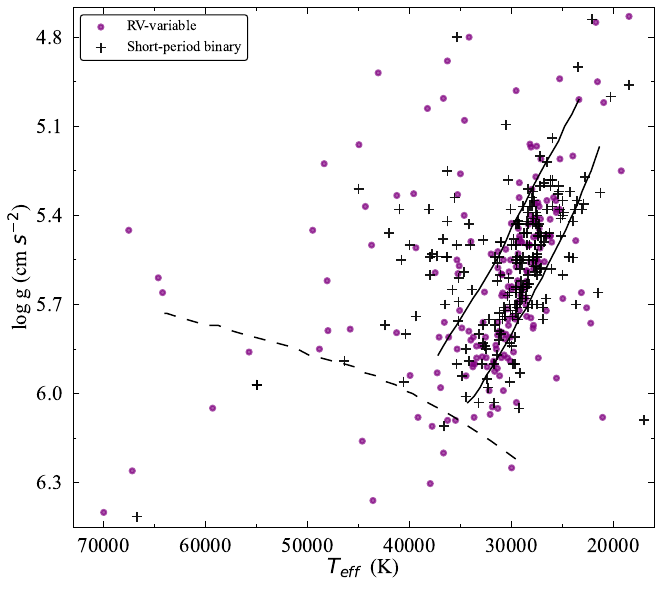}
      \caption{Distributions of RV-variable hot subdwarfs and the confirmed short-period binary hot subdwarfs from light curves and the literature in the $T_{\rm eff}$ -- $\log g$ diagram. The black crosses represent the confirmed short-period hot subdwarfs. The purple circles represent RV-variable hot subdwarfs.
              }
         \label{Fig.9}
         }
    \end{figure}

    In Fig. \ref{Fig.9} we combine the RV-variable hot subdwarfs (213) from our study and \citet{Geier.etal.2022} and compare their distributions with the confirmed short-period binary hot subdwarfs (210) from light curves and the literature. The two distributions are quite similar, except that more RV-variable stars are located below the He-MS. Only 5 $\pm$ 1\%, 6 $\pm$ 2\%, and 1 $\pm$ 1\% of the confirmed short-period binaries are found in the region in which EHBa, EHBc, and postEHBb stars are located, 
    which is consistent with the lower possibility of RV-variable stars to appear in these regions (7 $\pm$ 2\%, 7 $\pm$ 2\%, 6 $\pm$ 2\%). We found that 11 $\pm$ 2\% of the confirmed short-period binaries are grouped in the region in which post-EHBa stars are located. The value is similar to the fraction of 12 $\pm$ 2\% for RV-variable stars that appear in the region. As we mentioned before, these stars may have an evolutionary connection to EHBb stars. In addition, the simulations of \citet{Dorman.etal.1993} predicted that the EHB stars should spend most of their lifetimes during the He-shell burning stage at $T_{\rm eff}$ $\sim$ 35,000 -- 50,000 K, and then quickly evolve to the WD cooling stage. This means that these He-shell-burning stars which evolved from the EHB stars are more likely to be observed to cluster at the lower-temperature region above the EHB.
  
\section{Summary and conclusions}\label{sect.5}
  
   We have measured the RV variations of 317 single-lined and 117 composite hot subdwarfs through LAMOST DR11 LRS and MRS spectra. These stars were classified into different types according to their spectral line features and helium abundances to explore their RV-variability fractions. By combining our sample with that of \citet{Geier.etal.2022}, we studied the relation between the RV-variability of hot subdwarfs and their locations in the $T_{\rm eff}$ -- $\log g$ diagram. The light-variable hot subdwarfs searched through TESS light curves and the known short-period binary hot subdwarfs were also used to verify our results. The basic results are similar to those in \citet{Geier.etal.2022}, and we summarize them as follows. \\
   1. Only 6 $\pm$ 4\% of single-lined He-rich hot subdwarfs are found to be RV-variable. Their RV-variability fraction is significantly lower than that of single-lined He-poor sdB stars. Most of them may form through merger channels.  \\
   2. Single-lined sdB stars with $T_{\rm eff}$ $\sim$ 25,000 -- 33,000 K show an RV-variability fraction of 34 $\pm$ 5\%, whereas single-lined sdB stars cooler than about 25,000 k (11 $\pm$ 4\%) and single-lined sdB/OB stars with $T_{\rm eff}$ $\sim$ 33,000 -- 40,000 K and $\log g$ $\sim$ 5.7 -- 6.0 (13 $\pm$ 3\%) all exhibit a lower RV-variability fraction. \\
   3. The detected RV-variability fraction for single-lined hot subdwarfs that are located above the EHB band with $T_{\rm eff}$ $\sim$ 35,000 -- 45,000 K is 34 $\pm$ 9\%. The value is similar to the fraction of sdB stars at about 25,000 -- 33,000 K. These two subclasses of hot subdwarfs may have an evolutionary connection. However, single-lined hot subdwarfs stars with $T_{\rm eff}$ $\sim$ 45,000 -- 70,000 K show a lower RV-variability fraction of 10 $\pm$ 7\%. Some of them may not be connected to the evolution of sdB stars.\\
   4. Single-lined hot subdwarfs located below the canonical EHB show the largest RV-variability fraction of 51 $\pm$ 8\%. The majority of these stars may be low-mass hot subdwarfs or progenitors of He-WDs. \\
   5. The detected RV-variability fraction of composite hot subdwarfs (9 $\pm$ 3\%) is lower than the single-lined hot subdwarfs (25 $\pm$ 2\%). The stable RLOF channel may play an important role in the formation of composite hot subdwarfs.\\ 
   6. Eight composite sdB and two composite sdO stars were detected with RV variations larger than 45 km/s. These stars might be triple-system candidates or blends, and they require further observations to reveal their properties. \\

\section*{Data availability}

Table \ref{Table.A1} is only available in electronic form at the CDS via anonymous ftp to \url{cdsarc.u-strasbg.fr} (\url{130.79.128.5}) or via \url{http://cdsweb.u-strasbg.fr/cgi-bin/qcat?J/A+A/}.
 
\begin{acknowledgements}
   We thank the useful advice from Stephan Geier and Veronika Schaffenroth. This work is supported by the National Natural Science Foundation of China (Nos. 12288102 and 12333008) and National Key R$\&$D Program of China (No. 2021YFA1600403). X.M. acknowledges support from Yunnan Fundamental Research Projects (Nos. 202401BC070007 and 202201BC070003), International Centre of Supernovae, Yunnan Key Laboratory (No. 202302AN360001), the Yunnan Revitalization Talent Support Program Science $\&$ Technology Champion Project (NO. 202305AB350003), and the science research grants from the China Manned Space Project. Thanks to the LAMOST excellent spectra. The LAMOST Fellowship is supported by Special Funding for Advanced Users, budgeted and administered by the Center for Astronomical Mega-Science, Chinese Academy of Sciences (CAMS). Guoshoujing Telescope (the Large Sky Area Multi Object Fiber Spectroscopic Telescope LAMOST) is a National Major Scientific Project built by the Chinese Academy of Sciences. Funding for the project has been provided by the National Development and Reform Commission.
\end{acknowledgements}

\nocite{*}
\bibliography{references.bib}
\bibliographystyle{aa}

\begin{appendix}

\begin{table*}[h!]
\section{Additional material}
\caption{The RV variations and atmospheric parameters of 434 hot subdwarfs.}
\label{Table.A1}
\centering
\fontsize{9.5pt}{11pt}\selectfont
\begin{tabular}{ccccccccccc}
\hline\hline
\noalign{\smallskip}
R.A. & Decl. & $T_{\rm {eff}}$ & log g & $\log n{\rm (He)}/n{\rm (H)}$& Single-lined & Spclass & $\log p$ & Epoch & $\Delta RV_{\textup{max}}$ \\ 
(deg) & (deg) & (K) & (cm ${ \rm s^{-2}}$) & ~ & ~ & ~ & ~ & (Number)& (km/s)  \\
\noalign{\smallskip}
\hline
\noalign{\smallskip}
194.455489 & 54.42649 & 33130 $\pm$ 150 & 5.89 $\pm$ 0.02 & -1.55 $\pm$ 0.04 & is & sdOB & -0.001 & 9 & 3 $\pm$ 7 \\ 
257.555047 & 53.446121 & 24010 $\pm$ 440 & 5.54 $\pm$ 0.11 & -2.13 $\pm$ 0.11 & is & sdB & -14.78 & 2 & 89 $\pm$ 11 \\ 
56.022214 & 22.072953 & 28130 $\pm$ 440 & 5.59 $\pm$ 0.06 & -2.74 $\pm$ 0.09 & is & sdB & -3.02 & 7 & 40 $\pm$ 9 \\ 
71.237089 & 14.363909 & 32660 $\pm$ 320 & 5.66 $\pm$ 0.05 & -2.49 $\pm$ 0.13 & is & sdB & -7.84 & 2 & 64 $\pm$ 11 \\ 
98.497308 & 32.553966 & 26010 $\pm$ 1210 & 5.86 $\pm$ 0.11 & -2.97 $\pm$ 0.42 & is & sdB & -2.88 & 3 & 43 $\pm$ 16 \\ 
... &  & ... &  & ... &  & ... &  & ... &   \\
62.954352 & 15.383123 & 37490 $\pm$ 520 & 5.59 $\pm$ 0.04 & -3.23 $\pm$ 0.15 & is & sdO & -15.95 & 4 & 106 $\pm$ 16 \\
\noalign{\smallskip}
\hline
\noalign{\smallskip}
\noalign{\smallskip}
\multicolumn{11}{c}{
\begin{minipage}{\textwidth}
(This table is available in its entirety at the CDS. The uncertainty of $\Delta RV_{\textup{max}}$ is propagated from the uncertainties of its corresponding maximum and minimum RV.)
\end{minipage}
}
\end{tabular}
\end{table*}

\begin{table*}[h!]
\caption{The main parameters of our newly discovered short-period binary hot subdwarfs through light curves.}
\label{Table.A2}
\centering
\fontsize{9.5pt}{11pt}\selectfont
\begin{tabular}{ccccccccc}
\hline
\hline
\noalign{\smallskip}
    R.A. & Decl. & $T_{\rm {eff}}$ & log g & $\log n{\rm (He)}/n{\rm (H)}$ & Spclass & TIC & Type & Period \\ 
    (deg) & (deg) & (K) & (cm ${\rm s^{-2}}$) & ~ & ~ & (ID) & ~ & (h) \\ 
    \noalign{\smallskip}
    \hline 
    \noalign{\smallskip}
    346.401764 & 34.698363 & 28260 $\pm$ 160 & 5.50 $\pm$ 0.02 & -2.34 $\pm$ 0.04  & sdB & 369581468 & Reflection & 4.77 \\ 
    345.4409 & 13.64374 & 31190 $\pm$ 220 & 5.73 $\pm$ 0.05 & -1.74 $\pm$ 0.04 & sdB & 217587019 & Reflection & 3.92 \\ 
    312.413176 & 30.081818 & 37330 $\pm$ 170 & 5.54 $\pm$ 0.06 & -2.58 $\pm$ 0.12 & sdOB & 230775376 & Reflection & 10.31 \\ 
    4.23055 & 51.230486 & 32770 $\pm$ 460 & 5.48 $\pm$ 0.07 & -3.04 $\pm$ 0.58 & sdB & 202125132 & Reflection & 6.50 \\ 
    93.23016 & 57.847462 & 29270 $\pm$ 270 & 5.43 $\pm$ 0.03 & -2.28 $\pm$ 0.04 & sdB & 322550178 & Reflection & 3.09 \\ 
    118.30847 & 11.211171 & 29360 $\pm$ 60 & 5.42 $\pm$ 0.01 & -2.44 $\pm$ 0.04 & sdB & 468928859 & Reflection & 6.39 \\ 
    352.34432 & 32.233162 & 31070 $\pm$ 280 & 5.49 $\pm$ 0.07 & -2.59 $\pm$ 0.26 & sdB & 2054270826 & Reflection & 4.23 \\ 
    29.0050281 & 40.0561321 & 28380 $\pm$ 330 & 5.31 $\pm$ 0.03 & -2.62 $\pm$ 0.04 & sdB & 67423472 & Reflection & 4.65 \\ 
    207.719937 & 36.700612 & 66760 $\pm$ 2380 & 6.41 $\pm$ 0.07 & -1.07 $\pm$ 0.22  & sdO & 23709993 & Reflection & 79.55 \\ 
    77.542438 & 30.11271 & 29192 $\pm$ 263 & 5.93 $\pm$ 0.06 & -2.29 $\pm$ 0.09 & sdB & 367014246 & Reflection & 2.75 \\ 
    103.90969 & 31.394393 & 33232 $\pm$ 466 & 6.03 $\pm$ 0.11 & -1.96 $\pm$ 0.14 & sdOB & 741122759 & Reflection & 3.42 \\ 
    321.634188 & -4.223943 & 31412 $\pm$ 220 & 5.87 $\pm$ 0.05 & -2.82 $\pm$ 0.09 & sdB & 250262974 & Reflection & 4.48 \\ 
    109.74026 & 7.653687 & 45003 $\pm$ 731 & 5.31 $\pm$ 0.05 & -2.74 $\pm$ 0.13 & sdO & 264749962 & HW Vir & 2.03 \\ 
    114.89894 & 56.709241 & 29896 $\pm$ 553 & 5.90 $\pm$ 0.11 & -2.49 $\pm$ 0.12 & sdB & 742806233 & HW Vir & 8.20 \\ 
    316.00591 & 34.610072 & 30312 $\pm$ 644 & 5.28 $\pm$ 0.14 & -2.70 $\pm$ 0.07 & sdB & 1957912171 & HW Vir & 2.85 \\ 
    356.26688 & 50.965145 & 31748 $\pm$ 329 & 5.90 $\pm$ 0.09 & -2.63 $\pm$ 0.16 & sdB & 26994998 & HW Vir & 1.94 \\ 
    257.555047 & 53.446121 & 24010 $\pm$ 440 & 5.54 $\pm$ 0.11 & -2.13 $\pm$ 0.11 & sdB & 367779738 & Ellipsoidal & 1.82 \\ 
    288.815187 & 43.674379 & 27850 $\pm$ 240 & 5.57 $\pm$ 0.04 & -2.55 $\pm$ 0.04 & sdB & 158918567 & Ellipsoidal & 2.44 \\ 
    94.486139 & 18.83019 & 26600 $\pm$ 480 & 5.47 $\pm$ 0.05 & -2.91 $\pm$ 0.15 & sdB & 429807453 & Ellipsoidal & 2.06 \\ 
    181.601508 & 57.159922 & 35220 $\pm$ 350 & 5.76 $\pm$ 0.04 & -1.75 $\pm$ 0.06 & sdOB & 55753808 & Ellipsoidal & 9.98 \\ 
    8.943925 & 26.915094 & 27150 $\pm$ 520 & 5.58 $\pm$ 0.06 & -2.44 $\pm$ 0.09 & sdB & 301799840 & Ellipsoidal & 10.46 \\ 
    332.511298 & 25.066193 & 21290 $\pm$ 80 & 5.32 $\pm$ 0.02 & -2.39 $\pm$ 0.04 & sdB & 27782233 & Ellipsoidal & 4.00 \\ 
    317.098329 & 1.032085 & 25480 $\pm$ 500 & 5.33 $\pm$ 0.04 & -2.50 $\pm$ 0.06 & sdB & 387373406 & Ellipsoidal & 14.10 \\ 
    117.04874 & 13.730365 & 22940 $\pm$ 930 & 5.36 $\pm$ 0.16 & -2.68 $\pm$ 0.34 & sdB & 17561485 & Ellipsoidal & 3.76 \\ 
    91.999515 & 13.6144053 & 30580 $\pm$ 1510 & 5.09 $\pm$ 0.02 & -1.57 $\pm$ 0.12 & sdB & 151473446 & Ellipsoidal & 15.68 \\ 
    159.265844 & -0.138644 & 35357 $\pm$ 415 & 4.80 $\pm$ 0.04 & -2.51 $\pm$ 0.13 & sdB & 124598476 & Ellipsoidal & 44.56 \\ 
    238.9077349 & 27.1134627 & 22100 $\pm$ 900 & 4.74 $\pm$ 0.10 & -3.00 $\pm$ 0.10 & sdB & 258109545 & Ellipsoidal & 155.26 \\ 
    292.2527216 & 44.9497531 & 17001 $\pm$ 94 & 6.09 $\pm$ 0.02 & -2.43 $\pm$ 0.13 & sdB & 63208546 & Ellipsoidal & 7.10 \\ \hline
    \noalign{\smallskip}
\end{tabular}
\end{table*}

\end{appendix}
\end{document}